\documentclass[12pt]{article}

\usepackage{graphicx}

\usepackage[intlimits]{amsmath}
\usepackage{amssymb}
\usepackage{bbm}
\usepackage{mathrsfs}
\usepackage{wrapfig}
\usepackage{caption}
\usepackage{subfigure}
\usepackage{enumerate}
\usepackage{epsfig}
\usepackage{cite}

\def\lsim{\mathrel{\raise.3ex\hbox{$<$\kern-.75em\lower1ex\hbox{$\sim$}}}}
\def\gsim{\mathrel{\raise.3ex\hbox{$>$\kern-.75em\lower1ex\hbox{$\sim$}}}}

\def\T{{\bf 10}}
\def\F{{\bf 5}}

\def\1{{\nu^c}}

\def\mG{m_{\tilde{G}}}
\def\mg{m_{\tilde{g}}}
\def\mt{\widetilde{m}_1}
\def\eq{\mathrm{eq}}
\def\E{{\cal E}_{N_1}}

\def\n{\nu}

\def\bo{{\raise.15ex\hbox{\large$\Box$}}}               % D'Alembertian
\def\face{{\raise.2ex\hbox{$\displaystyle \bigodot$}\mskip-2.2mu \llap {$\ddot
        \smile$}}}                                      % happy face
\def\leftrightarrowfill{$\mathsurround=0pt \mathord\leftarrow \mkern-6mu
        \cleaders\hbox{$\mkern-2mu \mathord- \mkern-2mu$}\hfill
        \mkern-6mu \mathord\rightarrow$}       % <--> double differential
\def\dvec#1{\vbox{\ialign{##\crcr
        \leftrightarrowfill\crcr\noalign{\kern-1pt\nointerlineskip}
        $\hfil\displaystyle{#1}\hfil$\crcr}}}           % <--> accent
\def\beq{\begin{equation}}
\def\eeq{\end{equation}}

\def\beqx{\begin{displaymath}}
\def\eeqx{\end{displaymath}}

\def\beqa{\begin{eqnarray}}
\def\eeqa{\end{eqnarray}}

\catcode`@=11
\def\@citex[#1]#2{\if@filesw\immediate\write\@auxout{\string\citation{#2}}\fi
  \def\@citea{}\@cite{\@for\@citeb:=#2\do
    {\@citea\def\@citea{,\penalty\@m}\@ifundefined
      {b@\@citeb}{{\bf ?}\@warning
       {Citation `\@citeb' on page \thepage \space undefined}}%
\hbox{\csname b@\@citeb\endcsname}}}{#1}}

\def\citer{\@ifnextchar [{\@tempswatrue\@citexr}{\@tempswafalse\@citexr[]}}
\def\@citexr[#1]#2{\if@filesw\immediate\write\@auxout{\string\citation{#2}}\fi
  \def\@citea{}\@cite{\@for\@citeb:=#2\do
    {\@citea\def\@citea{--\penalty\@m}\@ifundefined
       {b@\@citeb}{{\bf ?}\@warning
       {Citation `\@citeb' on page \thepage \space undefined}}%
\hbox{\csname b@\@citeb\endcsname}}}{#1}}
\catcode`@=12
\begin{document}
\date{\mbox{ }}

\title{ 
{\normalsize     
DESY 10-128\hfill\mbox{}\\
August 2010\hfill\mbox{}\\}
\vspace{1cm}
\bf Matter and Dark Matter \\ from False Vacuum Decay \\[8mm]}
\author{W.~Buchm\"uller, K.~Schmitz, G. Vertongen\\[2mm]
{\normalsize\it Deutsches Elektronen-Synchrotron DESY, 22607 Hamburg, Germany}}
\maketitle

\thispagestyle{empty}

\vspace{1cm}
\begin{abstract}
\noindent
We study tachyonic preheating associated with the spontaneous breaking of
$B-L$, the difference of baryon and lepton number. Reheating occurs 
through the decays of heavy Majorana neutrinos
which are produced during preheating and in decays of the Higgs particles
of $B-L$ breaking. Baryogenesis is an interplay of nonthermal and thermal
leptogenesis, accompanied by thermally produced gravitino dark matter. 
The proposed mechanism 
simultaneously explains the generation of matter and dark matter, thereby
relating the absolute neutrino mass scale to the gravitino mass.
\end{abstract}

\newpage

\section{Introduction}

Thermal leptogenesis \cite{fy86} explains the observed matter-antimatter 
asymmetry of the universe in terms of neutrino masses that are consistent with
neutrino oscillation experiments. In its simplest version, the primordial
lepton asymmetry is generated by the $CP$ violating interactions of $N_1$, the
lightest of the heavy Majorana neutrinos, the seesaw partners of the
ordinary neutrinos. Typical values for the $N_1$ mass and for the leptogenesis
temperature are $M_1 \sim T_L \sim 10^{10}~\mathrm{GeV}$ \cite{bpy05}.

In supersymmetric theories such high reheating temperatures cause the
`gravitino problem' for heavy unstable gravitinos \cite{we82,ens85,kkm05}.
However, if the gravitino is the lightest superparticle (LSP), this problem can
become a virtue since for typical leptogenesis temperatures and for 
superparticle masses of the electroweak scale,
thermal production of gravitinos can explain the observed amount of
dark matter \cite{bbp98}. The dominant contribution from  QCD processes is
given by
\begin{align}
\Omega_{\tilde{G}} h^2 = C
\left(\frac{T_R}{10^{10}\,\textrm{GeV}}\right)
\left(\frac{100\,\textrm{GeV}}{m_{\tilde{G}}}\right)
\left(\frac{m_{\tilde{g}}}{1\,\textrm{TeV}}\right)^2 \ ,
\label{GDMestimate}
\end{align}
where $T_R$ is the reheating temperature while $m_{\tilde{G}}$ and 
$m_{\tilde{g}}$ are gravitino and gluino masses, respectively;  
the coefficient $C=0.27$ to leading order in the gauge coupling
\cite{bbb00,ps06}.\footnote{Note that $C$ has an $\mathcal{O}(1)$ uncertainty
due to unknown higher order contributions and nonperturbative effects 
\cite{bbb00}. Resummation of thermal masses increases $C$ by about a factor 
of two \cite{rs07}.} Hence, for $T_R \sim T_L$ and superparticle masses 
characteristic of gravity or gaugino mediation one indeed obtains the observed
amount of dark matter, $\Omega_{\mathrm{DM}} h^2 \simeq 0.11$ \cite{wmap10}.

Thermal leptogenesis is consistent with a reheating temperature $T_R$ much 
larger
than $T_L$, which is not the case for thermally produced gravitino dark matter.
Consistency of these two mechanisms for the origin of matter and dark matter
then requires an explanation of why $T_L$ and $T_R$ have the 
same order of magnitude. 

Successful leptogenesis, independent of initial
conditions, favours the mass window 
$10^{-3}~\mathrm{eV} \lsim m_i \lsim 0.1~\mathrm{eV}$ \cite{bdp03}
for the light neutrino masses . For
an effective neutrino mass 
$\widetilde{m}_1 = (m_D^\dagger m_D)_{11}/M_1 \sim 0.01~\mathrm{eV}$, where
$m_D$ is the Dirac neutrino mass, the decay width of the heavy Majorana 
neutrino $N_1$ becomes
\begin{equation}\label{mtilde}
\Gamma_{N_1}^0 = 
\frac{\widetilde{m}_1}{8 \pi} \left(\frac{M_1}{v_\textrm{\tiny EW}}\right)^2
\sim 10^3\,\textrm{GeV} \ ,
\end{equation}
where we have used $v_\textrm{\tiny EW} = 174~\mathrm{GeV}$ for the 
vacuum expectation value of electroweak symmetry breaking.
If the hot phase of the early universe is initiated by the decays of particles
with decay width $\Gamma$, the reheating temperature is given by 
\begin{equation}
T_R = \left(\frac{90}{8\pi^3 g_\ast}\right)^{1/4} \sqrt{\Gamma M_P} \ ,
\end{equation}
where $g_* \sim 200$ is the effective number of relativistic degrees of freedom
and $M_P = 1.22 \times 10^{19}~\textrm{GeV}$ is the Planck mass. If reheating
occurs through the decay of $N_1$ neutrinos, one obtains
\begin{align}
T_R \sim 0.2 \cdot \sqrt{\Gamma_{N_1}^0 M_P} \sim 10^{10}~\textrm{GeV} \ ,
\label{Treheat}
\end{align}
which is indeed the temperature needed for thermal leptogenesis and gravitino
dark matter.

This observation raises the question whether in the thermal bath produced
by heavy Majorana neutrino decays both, the matter-antimatter asymmetry and 
gravitino dark matter, can be simultaneously produced. In the following
we shall present an example which demonstrates that this is indeed possible.

\section{Flavour model and tachyonic preheating}

In supersymmetric extensions of the Standard Model the superpotential for 
matter fields including right-handed neutrinos reads
\begin{equation}\label{yuk}
W_M = h_{ij}^u \T_i\T_j H_u +  h_{ij}^d \F^*_i\T_j H_d
      + h_{ij}^{\n} \F^*_i n^c_j H_u 
      +  h_i^n n^c_i n^c_i S \ , 
\end{equation}
where we have used $SU(5)$ notation for ordinary quarks and leptons,
$i=1,2,3$ counts the generations and $n^c$ denote the charge conjugate
of right-handed neutrinos; $H_u$, $H_d$ and $S$ are the
electroweak and $B-L$ symmetry breaking fields, respectively,
\begin{align}
\langle H_u \rangle = v_u\ , \quad   \langle H_d \rangle = v_d\ , \quad 
\langle S \rangle = v_{B-L}\ .
\end{align}
The heavy Majorana neutrinos are given by $N_i = n_i + n^c_i$.

For the Yukawa couplings
we shall use a representative model \cite{by99} with a Froggatt-Nielsen $U(1)$ 
flavour symmetry, which is consistent with the observed hierarchical quark and 
lepton masses, thermal leptogenesis \cite{by99} as well as constraints from 
flavour
changing processes \cite{bdh99}. Their order of magnitude is given by
\begin{equation}\label{charges}
h_{ij} \propto \eta^{Q_i + Q_j}\ , 
\end{equation}
where the chiral $U(1)$ charges are listed in Table~\ref{tab:charges}; the
parameter $\eta \simeq 1/\sqrt{300}$ is determined by the hierarchy of quark 
and lepton masses. 

Using the seesaw mass relation and the estimate 
$\overline{m}_\n = \sqrt{m_3 m_2} \sim 0.01~\mathrm{eV}$, successful
thermal leptogenesis determines the mass of the lightest heavy Majorana
neutrino, $M_1 = 10^{10}~\mathrm{GeV}$ \cite{by99}. From
\begin{align}\label{N1mass}
M_1 = h^n_1  v_{B-L}  \simeq \eta^2  v_{B-L}  
\end{align} 
one then obtains $v_{B-L} \simeq 3\times 10^{12}~\mathrm{GeV}$.
The masses of the other heavy neutrinos coincide with the $B-L$
breaking scale,
\begin{align}
M_2 \simeq M_3 \simeq v_{B-L} \simeq 3\times 10^{12}~\mathrm{GeV}\ . 
\end{align}

\begin{table}[t]
\begin{center}
\begin{tabular}{c|cccccccccccccc}\hline \hline
$\psi_i$ & $\T_3$ & $\T_2$ & $\T_1$ & $\F^*_3$ & $\F^*_2$ & $\F^*_1$ &
$n^c_3$ & $n^c_2$ & $n^c_1$ & $H_u$ & $H_d$ & $S$  \\ \hline
$Q_i$ & 0 & 1 & 2 & 1 & 1 & 2 & 0 & 0 & 1 & 0 & 0 & 0  \\ \hline\hline
\end{tabular}
\medskip
\caption{Chiral $U(1)$ charges. $\T = (q,u^c,e^c)$, 
$\F^* = (d^c,l)$.} 
\label{tab:charges}
\end{center}
\end{table}

For simplicity, we shall only consider the Higgs boson of $B-L$ symmetry
breaking in the following\footnote{The supersymmetric Higgs mechanism for
$B-L$ breaking will be discussed in a forthcoming publication \cite{bsv10}.}.
The Higgs potential relates the $S$ mass
to the energy density of the unbroken phase,
\begin{align} 
m^2_S = \lambda v_{B-L}^2\ , \quad \rho_0 = \frac{1}{4}\lambda v_{B-L}^4\ .
\end{align}
We assume the Higgs coupling $\lambda$ to be ${\cal O}(1)$, such that the
Higgs mass lies in the range
\begin{align}
2 M_{2,3} > m_S > 2 M_1\ .
\end{align}

The $U(1)$ flavour symmetry also determines the decay rates of the
heavy Majorana neutrinos $N_1$, $N_{2,3}$ and of the Higgs boson $S$,
\begin{align}
\Gamma_{N_1}^0 &\simeq \frac{\eta^4}{8 \pi} M_1 \ , \label{N1width} \\
\Gamma_{N_{2,3}}^0 &\simeq \frac{\eta^2}{8 \pi} M_{2,3} \ , \label{N2width}\\
\Gamma_S^0 &= \frac{(h^n_1)^2}{16\pi} m_S
\left[1 - \left(2 M_1 / m_S\right)^2\right]^{3/2}
\simeq \frac{\eta^4}{16 \pi} m_S \ . \label{Swidth}
\end{align}
Note that these estimates for the decay widths have a considerable uncertainty
since the Yukawa couplings, which enter quadratically, are only known up
to factors $\mathcal{O}(1)$ \cite{by99}. 
For instance, since $M_1$ and $\Gamma_{N_1}^0$
involve different Yukawa couplings, the ratio $\Gamma_{N_1}^0/M_1$ can
easily vary by two orders of magnitude. Since this ratio will be important
in the following phenomenological analysis, we shall parametrize this
uncertainty explicitly by the effective neutrino mass $\mt$ introduced in
Eq.~(\ref{mtilde}).

A crucial ingredient of leptogenesis are the $CP$ asymmetries in heavy 
Majorana neutrinos decays \cite{crv96,bp98},
\begin{align}
\epsilon_i = \frac{1}{8\pi} \frac{1}{(h^{\n\dagger}h^{\n})_{ii}}
\sum_{j\neq i} \mathrm{Im}\left[(h^{\n\dagger}h^{\n})^2_{ij}\right]
F\left(\frac{M_j}{M_i}\right)\ ,
\end{align}
from which we obtain
\begin{align}
\epsilon_1 \sim 0.1 \ \eta^4\ \sim 10^{-6} , \quad   
\epsilon_{2,3} \simeq 0.1 \ \eta^2 \sim 3 \times 10^{-4} \ .
\end{align}
These asymmetries determine the lepton asymmetry which is generated by decays
and inverse decays of the heavy neutrinos. Note that for 
$M_1 \simeq 10^{10}~\mathrm{GeV}$, the estimate $\epsilon_1 \sim 10^{-6}$ 
corresponds to the maximal possible $CP$ asymmetry \cite{di02}. For smaller
heavy neutrino masses the asymmetry is reduced by the mass ratio,
$\epsilon_1 \sim 10^{-6}~M_1/10^{10}~\mathrm{GeV}$.

The vacuum energy of a phase with unbroken $B-L$ symmetry can drive a period
of inflation, which then rapidly ends due to the spinodal growth of long
wave-length modes, a process known as `tachyonic preheating' \cite{fgb01}.
The true vacuum is reached at a time $t_\textrm{PH}$ after inflation 
\cite{gbx01},
\begin{align}
\left.\langle S^{\dagger}S\rangle \right|_{t = t_\textrm{PH}} 
= v_{B-L}^2 \ , \qquad
t_\textrm{PH} \simeq \frac{1}{2m_S}
\ln\left(\frac{32 \pi^2}{\lambda}\right) \ .
\end{align}

During this phase transition the energy of the false vacuum is converted 
mostly into a nonrelativistic gas of $S$ bosons, with an admixture of
heavy neutrinos\footnote{For simplicity, we neglect the contribution of
scalar neutrinos ; also
in thermal leptogenesis their effect on the final
baryon asymmetry is known to be small \cite{plu98}.}
whose contribution to the energy density is determined by
their coupling to the Higgs field \cite{gbx01}, 
\begin{align}
r_{N_i} = \frac{\rho_{N_i}}{\rho_0} \simeq 1.5 \times 10^{-3} g_N
\lambda f\left(\alpha_i,0.8\right) \ ;
\label{eq:GarcBellRes}
\end{align}
here $g_N = 2$ and
$f\left(\alpha,\gamma\right) = \sqrt{\alpha^2 + \gamma^2} - \gamma$
with $\alpha_i = h^n_i / \sqrt{\lambda}$. 
For the neutrinos $N_{2,3}$ one obtains
\begin{align}
r_{N_{2,3}} \simeq 1 \times 10^{-3} (h^n_2)^2 
\simeq 1 \times 10^{-3} \ .  
\label{N2garbel}
\end{align}
The fraction $r_{N_1} = \mathcal{O}(\eta^4)$ and it is therefore negligible.

\section{Reheating, baryon asymmetry and gravitino dark matter 
from heavy neutrino decays}

We are now ready to study the time evolution of the initial state produced
by tachyonic preheating. First, the dominant contribution to the energy
density is the nonrelativistic gas of $S$ bosons (cf.~Eq.~(\ref{N2garbel})),
\begin{align}
\rho_S (t) = \frac{1}{a^3(t)} \left(1 - 2 r_{N_2}\right) \rho_0 \ ,
\end{align}
where $a(t)$ is the scale factor of the Friedman universe which is normalized 
by setting $a(t_\textrm{PH}) = 1$. It is convenient to introduce 
the comoving number densities for all particle species $X$,
\begin{align}
N_X(t) = a^3(t) n_X(t) \ .
\end{align}

At $t \sim t_2 = 1/\Gamma_{N_{2,3}}^0$ the
heavy neutrinos $N_2$ and $N_3$ decay, producing the initial radiation  
density\footnote{Since $t_2$ is very small compared to the time scales
relevant for leptogenesis, we approximate the onset of radiation by
a step function.}
\begin{align}
\rho_R (t_2) = \frac{2 r_{N_2}\rho_0}{a^3(t_2)} \ ,
\end{align}
with temperature
\begin{align}
T\left(t_2\right) = \left(\frac{30}{\pi^2 g_*} 
\frac{2 r_{N_2}\rho_0}{a^3(t_2)}\right)^{1/4} \ .
\end{align}
The out-of-equilibrium decay of $N_2$ and $N_3$ also produces a $B-L$
asymmetry\footnote{In \cite{gbx01} only the initial $B-L$ asymmetry 
from tachyonic preheating is
taken into account.}. The corresponding comoving number density is given by
\begin{align}
N_{B-L} \left(t_2\right) = \epsilon_2 N_{N_2} \left(t_2\right)
+ \epsilon_3 N_{N_3} \left(t_2\right)
\simeq 0.2~\eta^2 N_{N_2}(t_\textrm{PH}) \ .
\end{align}

The scale factor $a(t)$ has to be determined by solving the Friedman equation.
For a flat universe and
constant equation of state $\omega = p/\rho$ between some initial time 
$t_0$ and time $t$, one has 
\begin{align}
a(t) = a(t_0)\left[1 + \frac{3}{2}\left(1 + \omega\right)
\left(\frac{8 \pi}{3 M_P^2}\rho_\textrm{tot}\left(t_0\right)\right)^{1/2}
\left(t - t_0\right)\right]^{\frac{2}{3\left(1+\omega\right)}} \ .
\end{align}
One expects that until the decay of the $S$ bosons at 
$t \sim t_S = 1/\Gamma_S^0$ the system is mostly nonrelativistic and that
it becomes relativistic at later times. We have checked numerically that
this is indeed the case with effective equation-of-state parameters
$3(1+\omega_\textrm{PH}) \simeq 3.3$ for $t_\textrm{PH} < t \leq t_S$ and 
$3(1+\omega_S) \simeq 4.0$ for $t > t_S$, respectively.

The thermal part of the plasma due to $N_2$ and $N_3$ decays produces also
$N_1$ neutrinos. Their comoving number density $N_1^T$ satisfies the 
familiar Boltzmann equation
\begin{align}
a H \frac{d}{da} N_{N_1}^T = 
- \Gamma_{N_1} \left(N_{N_1}^T - N_{N_1}^\eq\right) \ ; \label{BE1}
\end{align}
here $H = \dot{a}/a$ is the Hubble parameter, and the thermal width
\begin{align}
\Gamma_{N_1} = \Gamma^0_{N_1} \frac{K_1(z)}{K_2(z)} \ , 
\quad z=\frac{M_1}{T} \ ,
\end{align}
where $K_1$ and $K_2$ are modified Bessel functions. Note that in deriving
Eq.~(\ref{BE1}) one assumes kinetic equilibrium for the thermally
produced $N_1$ neutrinos.

$N_1$ neutrinos are also produced in $S$ decays. These neutrinos are 
relativistic, but not in kinetic equilibrium. They are produced with energy
$m_S/2$ which is then redshifted with increasing scale factor. We take this
into account by solving the Boltzmann equations for the $S$ 
and $N_1$ distribution functions ($E_{N_1} = \sqrt{p^2 + M_1^2}$),
\begin{align}
\left(\frac{\partial}{\partial t} 
- H p \frac{\partial}{\partial p}\right) f_S(t,p) = 
&- \frac{m_S}{E_S} \Gamma_S^0 f_S(t,p) \ ,  \\
\left(\frac{\partial}{\partial t} 
- H p \frac{\partial}{\partial p}\right) f_{N_1}^S(t,p) = 
&-\frac{M_1}{E_{N_1}} \Gamma_{N_1}^0 f_{N_1}^S(t,p) \nonumber\\
&+ \frac{2\pi^2 n_S \Gamma_S^0}{E_{N_1}^2}
\left[1 - \left(2M_1 / m_S\right)^2\right]^{-1/2}
\delta\left(E_{N_1} - m_S /2\right) \ . 
\end{align}
A straightforward calculation yields for the distribution functions
\begin{align}
f_S(t,p) = & 
\frac{2\pi^2 N_S(t_2)}{k^2} \delta (k) 
e^{-\Gamma_S^0\left(t-t_2\right)}\ ,\quad p=\frac{k}{a(t)}\ , \label{fSdis} \\
f_{N_1}^S(t,p) = &
\frac{1}{a^3} \frac{4\pi^2 \Gamma_S^0}{m_S}
\left(1 - \left(2M_1 / m_S\right)^2\right)^{-\frac{1}{2}} 
\int_{t_2}^t dt' \Bigg[\frac{a}{a'}
\delta\left(E_{N_1} - \E\left(t',t\right)\right) \nonumber \\
&\times \E^{-1}\left(t',t\right) N_S\left(t'\right)
\exp\left(-M_1 \Gamma_{N_1}^0 \int_{t'}^t dt''
\E^{-1}\left(t',t''\right)\right)\Bigg] \ , 
\end{align}
where $\E(t',t)$ is the redshifted energy at time $t$ of an $N_1$ neutrino
produced in $S$ decay at time $t'$ with energy $m_S/2$,
\begin{align}
\E\left(t',t\right) = \frac{m_S}{2}\frac{a'}{a}
\left[1 + \left(\left(\frac{a}{a'}\right)^2 - 1\right)
\left(\frac{2M_1}{m_S}\right)^2\right]^{\frac{1}{2}} \ .
\end{align}
The corresponding comoving number densities are easily obtained from
\begin{align}
N_X(t) = a^3 g_X \int \frac{d^3p}{(2\pi)^3} f_X(t,p)\ ,
\end{align}
where $X=S,N_1$.

The number densities of thermally and nonthermally produced $N_1$ neutrinos
enter as source terms in the Boltzmann equation for the $B-L$ 
asymmetry, which reads
\begin{align}
a H \frac{d}{da} N_{B-L} = \epsilon_1 \Gamma_{N_1} \left(N_{N_1}^T
- N_{N_1}^\eq\right) - \frac{N_{N_1}^\eq}{2 N_{L}^\eq}\Gamma_{N_1} N_{B-L}
+ \epsilon_1 \Gamma_{N_1}^0 \widetilde{N}_{N_1}^S \ , \label{BE2}
\end{align}
with the nonthermal contribution
\begin{align}
\widetilde{N}_{N_1}^S (t)= \int_{t_2}^t dt' 
\frac{dN_{N_1}^S(t')}{dt'}\frac{M_1}{\E(t',t)} \ , 
\end{align}
which includes the relativistic correction factor for the decays of $N_1$
as function of time.

In addition to the baryon asymmetry we are interested in thermal production
of gravitinos which is governed by the Boltzmann equation
\begin{align}
a H \frac{d}{da} N_{\tilde{G}} = a^3 {\cal C}_{\tilde{G}}(T) \ , \label{BE3}
\end{align}
with the dominant QCD collision term in the supersymmetric Standard Model 
\cite{bbb00,ps06}
\begin{align}
{\cal C}_{\tilde{G}}(T) = \left(1 + \frac{m_{\tilde{g}}^2}{3 \mG^2}\right)
\frac{54 \zeta(3) g_s(T)^2}{\pi^2 M_P} T^6
\left[\ln\left(\frac{T^2}{m_g(T)^2}\right) + 0.8846\right] \ ;
\end{align}
here $g_s(10^{10}~\mathrm{GeV}) = 0.85$ is the QCD gauge coupling, 
and $m_g(T) = \sqrt{3/2}g_s(T)T$ is the plasma mass of the gluon.

Finally, we need the temperature as function of the scale factor in order
to compute the collision term for gravitino production and the equilibrium
number density of the $N_1$ neutrinos. The covariant energy conservation
connects the thermal contributions to energy density and pressure with the
nonthermal contributions,
\begin{align} 
\frac{d}{dt}\left(\rho_R + \rho_{N_1}^T + \rho_S + \rho_{N_1}^S\right)  
+ 3H\left(\rho_R + \rho_{N_1}^T + \rho_S + \rho_{N_1}^S 
+ p_R + p_{N_1}^T + p_{N_1}^S\right) = 0 \ .
\end{align}
The quantities $\rho_S$, $\rho_{N_1}^S$ and $p_{N_1}^S$ are easily obtained
from the corresponding distribution functions; energy and pressure of the
thermally produced $N_1$ neutrinos are given in terms of the number density,
\begin{align}
a^3 p_{N_1}^T = N_{N_1}^T T \ , \quad 
a^3 \rho_{N_1}^T = T^2 \frac{\partial}{\partial T} N_{N_1}^T \ .
\end{align}
Together with the Boltzmann equations (\ref{BE1}), (\ref{BE2}) and (\ref{BE3}) 
we now have a complete
system of first-order differential equations which determine the time 
evolution of the initial state, in particular the generation of 
baryon asymmetry and gravitino abundance.

\section{Results and discussion}

We have numerically solved the Boltzmann equations derived in the previous
section. In the following we describe the solution for a representative
choice of neutrino masses, $CP$ asymmetries\footnote{We have chosen opposite
signs for the $CP$ asymmetries $\epsilon_1$ and $\epsilon_{2,3}$, so that
one can distinguish their contribution to the final $B-L$ asymmetry.}
and superparticle masses: 
$\mt = 10^{-3}~\mathrm{eV}$, $M_1 = 10^{10}~\mathrm{GeV}$, 
$\epsilon_1 = 10^{-6}$, $\epsilon_{2,3} = -3\times 10^{-4}$, and
$\mG = 100~\mathrm{GeV}$, $\mg = 800~\mathrm{GeV}$. 
Technical details of the solution and a systematic study
of the parameter space compatible with leptogenesis and gravitino dark
matter will be presented in a forthcoming publication \cite{bsv10}.

The various components of the energy density and the comoving 
number densities are compared as functions of the scale factor $a$ 
in Fig.~1 and Fig.~2,
respectively. Immediately after the end of tachyonic preheating
the decay of $N_2$ and $N_3$ generates radiation, an initial $B-L$ 
asymmetry and subsequently
a thermal abundance of $N_1$ neutrinos. At $a \sim 2$, the 
$S$ bosons decay. The produced $N_1$ neutrinos are relativistic, but 
nonthermal. At $a \sim 150$, the thermal $N_1$ neutrinos reach thermal 
equilibrium, the initial $B-L$ asymmetry is washed out, a new $B-L$ asymmetry
and the dominant part of radiation are produced in decays of the 
nonthermal $N_1^S$ neutrinos. Furthermore, gravitinos are continuously
produced from the thermal bath. Around $a \sim 1000$ the final $B-L$ asymmetry 
and the gravitino number density are reached.

\begin{figure}
\begin{center}
\epsfig{file=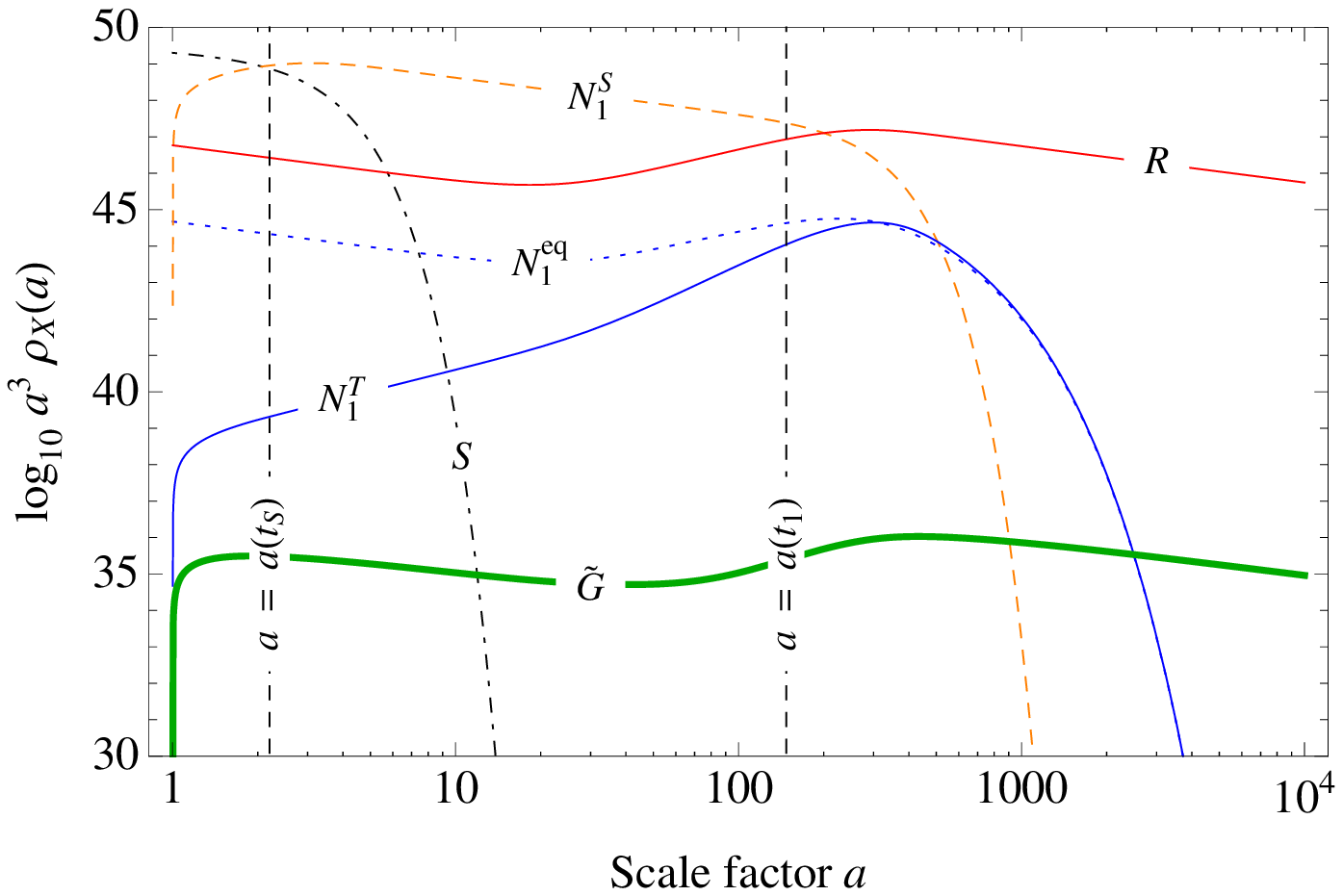,width=13cm}
\caption{Comoving energy densities for $S$ bosons, radiation ($R$),
$N_1$ neutrinos produced in $S$ decays ($N_1^S$),
$N_1$ neutrinos in equilibrium ($N_1^{\eq}$, for comparison), 
thermally produced 
$N_1$ neutrinos ($N_1^T$) and gravitinos ($\tilde{G}$) as functions of
the scale factor $a$.}
\vspace{5mm}
\epsfig{file=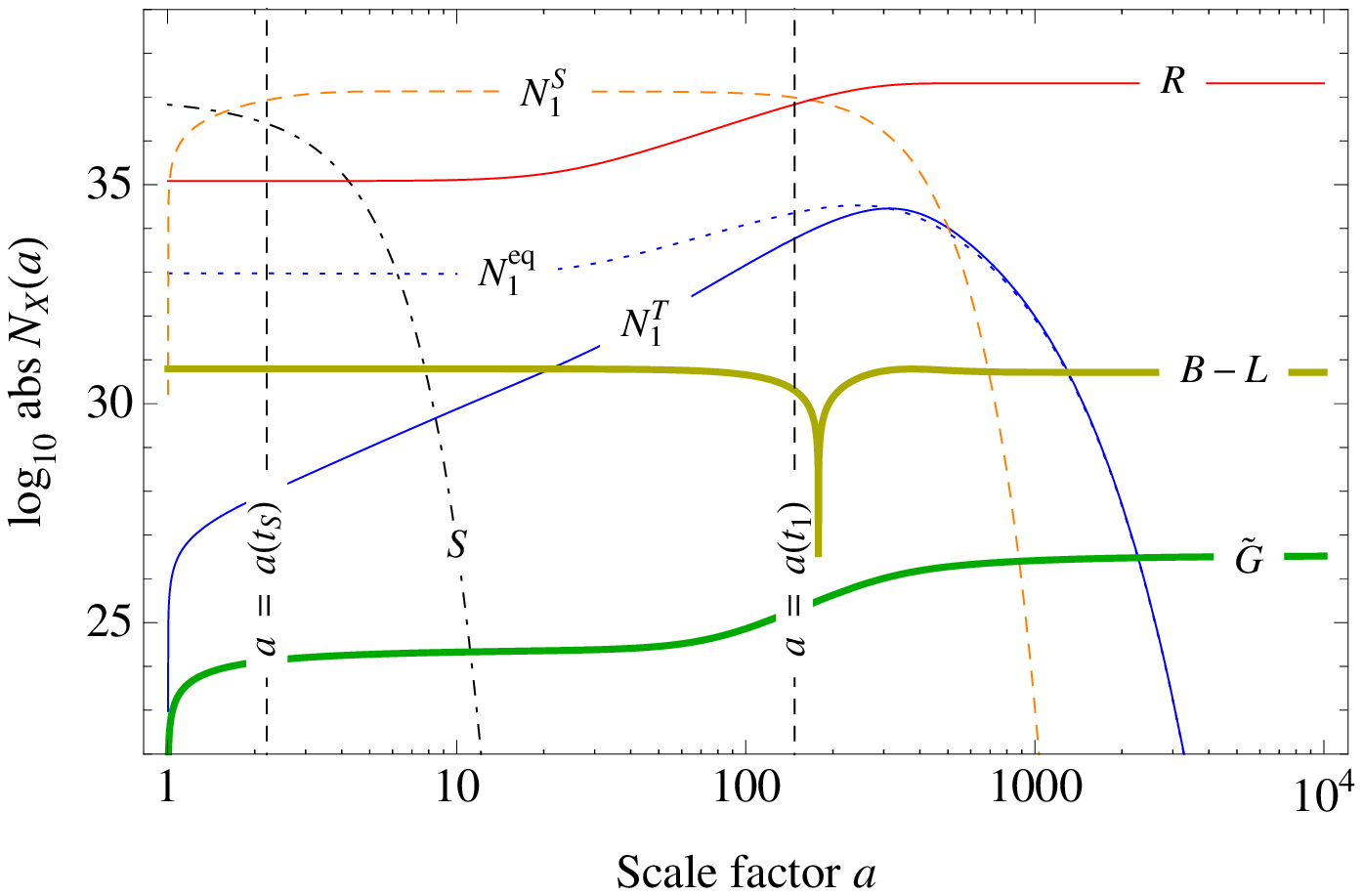,width=13cm}
\caption{Comoving number densities for $S$ bosons, $B-L$ charge ($B-L$) 
$N_1$ neutrinos produced in $S$ decays ($N_1^S$),
$N_1$ neutrinos in equilibrium ($N_1^{\eq}$, for comparison), 
thermally produced 
$N_1$ neutrinos ($N_1^T$) and gravitinos ($\tilde{G}$) as functions of
the scale factor $a$.}
\end{center}
\end{figure}

The present value of baryon asymmetry is obtained from
\begin{align}
\eta_B = \frac{n_B^0}{n_{\gamma}^0} 
= c_{\mathrm{sph}} \frac{g_*^0}{g_*}\frac{N_{B-L}}{N_{\gamma}}\Big|_{a_f} \ ,
\end{align}
where $a_f \sim 10^4$ is a scale factor after leptogenesis is completed; in
the supersymmetric Standard Model the sphaleron conversion factor 
$c_{\mathrm{sph}} = 8/23$, the effective numbers of degrees of freedom at
high and low temperatures are $g_* = 915/4$ and $g_*^0 = 43/11$, respectively,
and the number density of photons is 
$N_{\gamma} = a^3 g_{\gamma}\zeta{(3)}/\pi^2 T^3$. 
The corresponding expression for
the present gravitino abundance is given by
\begin{align}
\Omega_{\tilde{G}} = \frac{\mG n_{\gamma}^0}{\rho^0_c} \frac{g_*^0}{g_*}
\frac{N_{\tilde{G}}}{N_{\gamma}}\Big|_{a_f} \ ,
\end{align}
where $\rho^0_c = 1.052\times 10^{-5}~h^2~\mathrm{GeV}\mathrm{cm}^{-3}$
is the critical density. For our choice of parameters we obtain
\begin{align}\label{result}
\eta_B = 1.6 \times 10^{-7}\ , \quad \Omega_{\tilde{G}} h^2 = 0.11 \ .
\end{align}
The gravitino abundance is close to the observed value for 
$\Omega_{\mathrm{DM}}$. The calculated baryon asymmetry is about two orders
of magnitude larger than the observed one,  
$\eta_B^{\mathrm{CMB}} \simeq 6.2 \times 10^{-10}$ \cite{wmap10}, which is
consistent since we have used the maximal $CP$ asymmetry.

\begin{figure}
\begin{center}
\epsfig{file=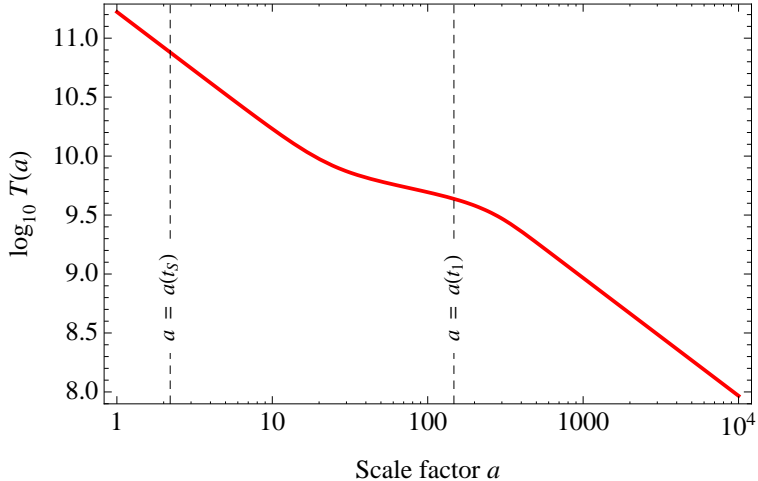,width=10cm}
\caption{Radiation temperature as function of the scale factor $a$.}
\end{center}
\end{figure}

The temperature of radiation is shown in Fig.~3 as function of the scale 
factor. It 
decreases like $T \propto 1/a$ except for an approximate plateau around 
$T_L \simeq 5\times 10^9~\mathrm{GeV}$ at $a \sim 100$, when the 
$N_1$ neutrinos decay. Note that the temperature $T_L$ agrees with
the reheating temperature $T_R$ estimated in Eq.~(\ref{Treheat})
up to a factor of two.
Inserting the temperature $T_L$ in the estimate Eq.~(\ref{GDMestimate})
for the gravitino abundance, one obtains agreement with the result  of the
full calculation given in Eq.~(\ref{result}). For such large reheating
temperatures nonthermal production of gravitinos is usually negligible
\cite{nty10}.

It is instructive to compare the calculated baryon asymmetry with two
extreme cases, namely thermal leptogenesis and the rapid conversion of a
gas of nonrelativistic $N_1$ neutrinos dominating the energy density of
the universe. For thermal leptogenesis one obtains for
our choice of parameters (cf.~\cite{bdp03})
\begin{align}
\eta_B^{\mathrm{thermal}} =  
\frac{3}{4} \frac{g_*^0}{g_*} c_{\mathrm{sph}} \epsilon_1 \kappa_f (\mt)
\simeq 5\times 10^{-10}\ ,
\end{align}
where we have used $c_{\mathrm{sph}} = 8/23$ and 
$\kappa_f (10^{-3}~\mathrm{eV}) \simeq 0.1$ for the
efficiency factor.
In the case of rapid conversion, energy conservation yields (cf.~\cite{kt90})
\begin{align}
\eta_B^{\mathrm{rapid}}  
\simeq 7 \frac{3}{4} c_{\mathrm{sph}} \epsilon_1\frac{T_L}{M_1}
\simeq 9\times 10^{-7} \ ,
\end{align}
which is closer to the result of our calculation Eq.~(\ref{result}). Varying
the parameters $\mt$ and $M_1$ one can interpolate between thermal
and nonthermal leptogenesis. The effective neutrino mass $\mt$ is closely
related to the smallest neutrino mass $m_1$ and therefore to the
absolute neutrino mass scale.
Via its effect on the reheating temperature and
gravitino dark matter, the absolute neutrino mass scale is thus related
to the gravitino mass.

The case of rapid conversion is also realized in leptogenesis from inflaton 
decays \cite{ahx99}. Here the reheating temperature is determined by the 
decay width of the inflaton. One can then have $T_R \ll M_1$ and in this
way avoid overproduction of gravitinos. For larger inflaton decay widths 
reheating temperatures $T_R \sim M_1$ can be reached, associated with
larger baryon asymmetries \cite{hwp08}, similar to the situation described 
in this paper.

We have shown that reheating the universe by heavy Majorana neutrino
decays can simultaneously explain the cosmological baryon asymmetry and
the observed dark matter abundance in terms of thermally produced gravitinos.
Starting from an initial state of unbroken $B-L$ symmetry, tachyonic
preheating leads to an interplay of nonthermal and thermal leptogenesis,
which is controlled by the heavy neutrino decay widths. 
Open questions concern the connection with models of inflation and
tachyonic preheating in supersymmetric theories.

\end{document}